# $SnP_2S_6$: A Promising Infrared Nonlinear Optical Crystal with Strong Non-Resonant Second Harmonic Generation and Phase-matchability

*Jingyang He[†], Seng Huat Lee[§ ‖], Francesco Naccarato[‡], Guillaume Brunin[‡], Rui Zu[†], Yuanxi Wang[§ ^], Leixin Miao[†], Huaiyu Wang[†], Nasim Alem[†], Geoffroy Hautier[‡ ⊥], Gian-Marco Rignanese[‡], Zhiqiang Mao[§ ‖], Venkatraman Gopalan[† ‖]\**

[†]Department of Materials Science and Engineering, Pennsylvania State University, University Park, Pennsylvania, 16802, USA

[§]2D Crystal Consortium, Materials Research Institute, Pennsylvania State University, University Park, Pennsylvania, 16802, USA

[‖]Department of Physics, Pennsylvania State University, University Park, Pennsylvania, 16802, USA

[‡]Institute of Condensed Matter and Nanosciences, UCLouvain, 1348 Louvain-la-Neuve, Belgium

[⊥]Thayer School of Engineering, Dartmouth College, Hanover, NH, 03755, USA

[^] Department of Physics, University of North Texas, Denton, Texas, 76203, USA

ABSTRACT

High-power infrared laser systems with broadband tunability are of great importance due to their wide range of applications in spectroscopy and free-space communications. These systems require nonlinear optical (NLO) crystals for wavelength up/down conversion using sum/difference frequency generation, respectively. NLO crystals need to satisfy many competing criteria, including large nonlinear optical susceptibility, large laser induced damage threshold (LIDT), wide transparency range and phase-matchability. Here, we report bulk single crystals of $SnP_2S_6$ with a large non-resonant SHG coefficient of $d_{33}$ = 53 pm $V^{-1}$ at 1550nm and a large LIDT of 350 GW $cm^{-2}$ for femtosecond laser pulses. It also exhibits a broad transparency range from 0.54μm to 8.5μm (bandgap of ~2.3 eV) and can be both Type I and Type II phase-matched. The complete linear and SHG tensors are measured as well as predicted by first principles calculations, and they are in excellent agreement. A proximate double-resonance condition in the electronic band structure for both the fundamental and the SHG light is shown to enhance the non-resonant SHG response. Therefore, $SnP_2S_6$ is an outstanding candidate for infrared laser applications.

**Introduction**

Nonlinear optical (NLO) crystals are essential components in high-power laser systems. They can produce coherent laser radiation at the wavelength of interest by combining or splitting photons, called sum (SFG) and difference (DFG) frequency generation.[1] Second harmonic generation (SHG) is an SFG process by which two identical photons combine into one at twice the frequency.[1,2] Over the past decades, there has been increasing interest in discovering new infrared NLO crystals because of their scientific and technological applications such as remote sensing,[3] biomedical imaging[4] and medical surgery.[5] Traditional NLO materials such as β-$BaB_2O_4$, $LiB_3O_5$[6] and LiNbO$_3$[7] are of limited usage in the mid-infrared regime due to strong multiphonon absorption. Currently, the commercially available infrared NLO crystals are AgGaS$_2$,[8] AgGaSe$_2$[9] and ZnGeP$_2$.[9,10] They exhibit large NLO coefficients and broad transmission range; however, they also suffer some intrinsic drawbacks such as rather low laser induced damage threshold (LIDT)[11,12] and strong absorption of 1-2μm,[13] limiting the choice of pump lasers.[10,14] To qualify as a good NLO material, there are several criteria: (1) large SHG coefficients to have efficient NLO conversion; (2) wide transparency window allowing broadband tunability; (3) large LIDT and (4) phase-matchability for good SHG efficiency.[10,15–17] It is well known that the main challenge in the discovery and optimization of NLO materials is the tradeoff between the bandgap and the magnitude of the NLO coefficients: materials with larger bandgaps typically have lower SHG coefficients but higher LIDTs.[17,18] Finding new infrared NLO crystals with a superior optimized set of optical properties to the current materials is a grand challenge in this field.

Metal thiophosphates are a promising compound system for NLO applications in the infrared region.[15] The covalent characteristics of the P-S bonds not only give rise to a large NLO response but also lead to larger bandgaps (possibly higher LIDT value),[15] both of which are essential for practical applications. Sn$_2$P$_2$S$_6$ is one of the metal thiophosphates whose linear and nonlinear optical properties are well studied.[19,20] Though it has a wide transparency window up to 8 μm and can be phase-matched, it has a relatively small SHG coefficient of $d_{11}$~17 pm V$^{-1}$.[20] SnP$_2$S$_6$, closely related to Sn$_2$P$_2$S$_6$, is a promising NLO material.[21] It was first synthesized in 1995 and has been reported to demonstrate excellent SHG response in powder form;[21,22] however, its detailed optical properties have not been studied due to the lack of large, high-quality single crystals. Recently, Zhang et al. reported the SHG coefficient in two-dimensional SnP$_2$S$_6$ at a wavelength of 810nm;[23] however, since the crystal has a bandgap of ~2.3eV, the measured SHG at 405nm is resonant in this case. For practical laser applications, the material must demonstrate large *non-resonant* nonlinearity for all photon energies below the bandgap. To address this question, we studied the experimental and theoretical linear and *non-resonant* nonlinear optical susceptibilities to evaluate its potential as a future NLO bulk material. We report that bulk single crystals of $SnP_2S_6$ demonstrate large SHG coefficients of *d*$_{33}$=53 pm V$^{-1}$ at 1550nm, surpassing $AgGaS_2$ (13.7 pm V$^{-1}$)[24] and $AgGaSe_2$ (33 pm V$^{-1}$).[25] $SnP_2S_6$ is also determined to be both Type I and Type II phase-matched at 1550nm with $d_{eff,I}$ ≈20 pm V$^{-1}$ and $d_{eff,II}$ ≈15 pm V$^{-1}$. It also exhibits a broad transparency range from 0.54μm to 8.5μm (bandgap of ~2.3 eV). First principles theory sheds light the crystal chemistry reasons for the superior properties, especially the double-near-resonance condition for the SHG arising from the electronic band structure. Quantitative measurements and theory predictions of the complete linear and SHG tensors are in excellent agreement. These excellent properties make it an outstanding candidate as a bulk NLO crystal for next-generation infrared laser systems.

**Results and Discussion**

The single crystals of $SnP_2S_6$ were grown by chemical vapor transport using the method reported by Wang et al.[21] It crystallizes in the space group of *R*3 and consists of a chiral and layered atomic arrangement, as shown in Figure 1a. In each layer, the P atoms are bonded covalently with each other and connect the P-S sublayer, in which the P-S bonds exhibit both covalent and ionic characteristics. The crystal structure of $SnP_2S_6$ is closely related to the parent compound $Sn_2P_2S_6$ except that half of the $Sn^{2+}$ ions are replaced with $Sn^{4+}$ ions, and the other half are vacant. Figure 1b shows the atomic resolution high angle annular dark-field scanning transmission electron microscopy (HAADF-STEM) image taken along the *a*-axis. The STEM image confirmed that the layered structure of the as-grown $SnP_2S_6$ crystals consisted of the stacking of S-P-P-S layers, which agreed well with structures previously reported (Figure 1a).[21] X-ray diffraction (XRD) collected on a single crystal SnP$_2$S$_6$ indicates its phase purity and high crystallinity, as shown in Figure 1c. The out-of-plane lattice parameter $c$ extracted from the XRD (000$l$) pattern is 19.1Å, consistent with the previously reported value of 19.424Å.[21]

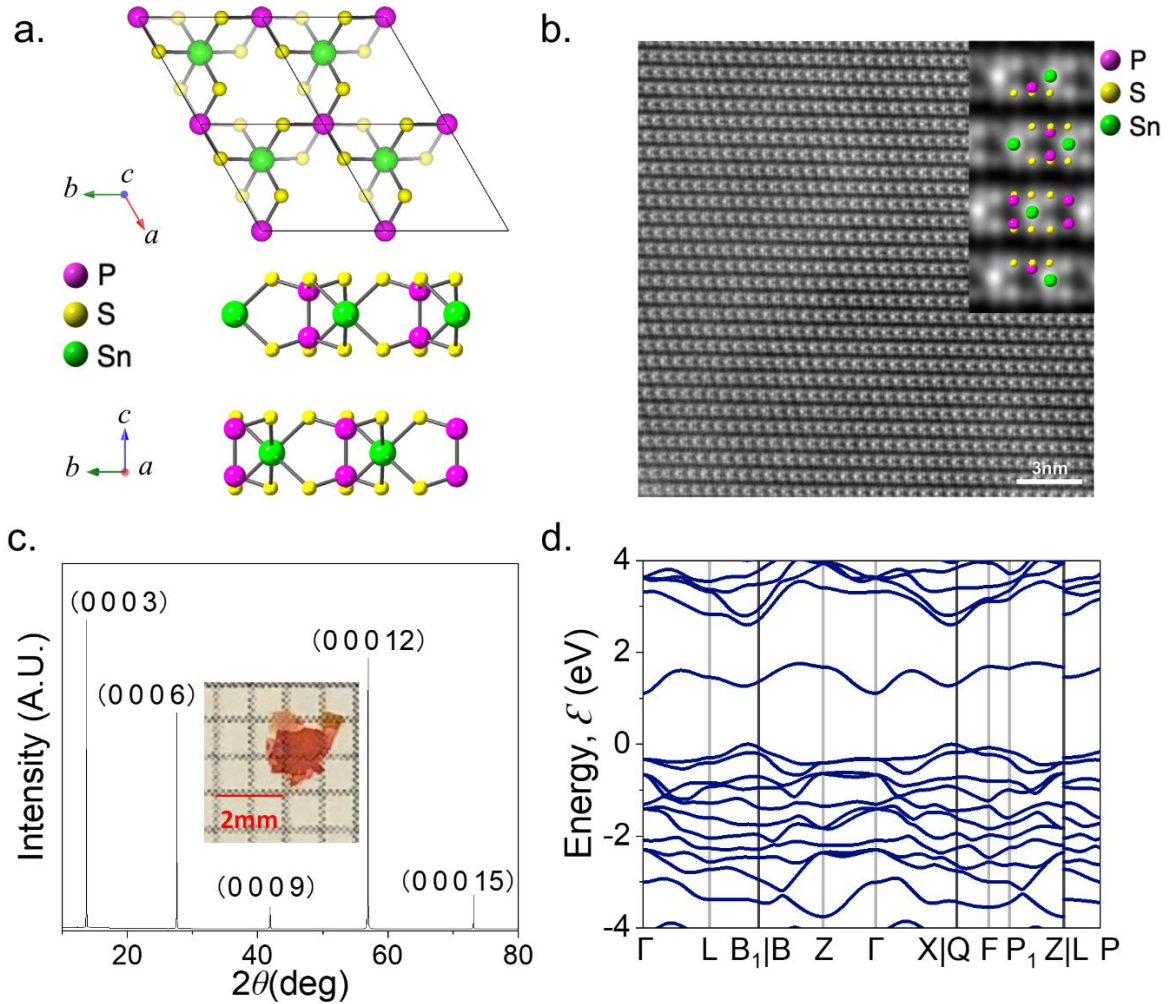

**Figure 1.** (a) Top view and side view of the SnP$_2$S$_6$ layers. (b) STEM image taken from the $a$-axis. Inset: A magnified HAADF-STEM image superimposed with a unit cell indicates the layered structure. (c) XRD pattern of a high-quality SnP$_2$S$_6$ crystal. All peaks match well with the (000$l$) plane of the bulk crystal, and the $c$ lattice constant is calculated to be ~19.1Å which is consistent

with the reported value of $c = 19.424$ Å. (d) Computed band structure of $SnP_2S_6$ for the LDA structure.

First-principles (FP) calculations were additionally performed to characterize the crystal structure of $SnP_2S_6$. After structural relaxation with the local density approximation (LDA), the calculated cell parameters are $a = 5.919$ Å and $c = 18.803$ Å. The former is in very good agreement (difference < 2%) with the previously reported experimental lattice parameter ($a = 5.999$ Å), while the discrepancy is slightly larger for the latter (3.2%). For this reason, the unit cell was also relaxed including van der Waals (vdW) interactions, leading to $a = 5.985$ Å and $c = 19.460$ Å, in excellent agreement with the experimental values. FP calculations were also carried out to determine the band structure of $SnP_2S_6$, given in Figure 1d. As expected within LDA, the bandgap is severely underestimated. For the LDA structure, the fundamental gap is 1.1 eV and the direct gap is 1.3 eV. The first conduction band consists of an isolated electronic band with a bandwidth of 0.7 eV. For the vdW structure, the band structure remains similar (Figure S1), with fundamental and direct gaps of 1 and 1.1 eV, respectively, and the first conduction band has a similar bandwidth.

We next investigated the linear optical properties of $SnP_2S_6$. The transparency window of $SnP_2S_6$ was measured in transmission through a sample with the (0001) surfaces by Ultraviolet-Visible-near-IR Spectroscopy (UV-Vis-NIR) (over 0.45μm to 2μm) and Fourier transform infrared (FTIR) spectroscopy (over 1μm to 20μm). During the FTIR measurements, a 15×objective was used to focus the beam on the area with uniform thickness. Thus, interference patterns were observed when multiple reflections occurred inside the sample and caused constructive/destructive interference. This was later used to determine the refractive index. No interference was detected in the UV-Vis-NIR data owing to the larger beam size used (beam diameter of 2mm) and the nonuniform thickness of the sample. The artifacts at ~0.86 μm are due to a change in the detector. From the transmittance (*T*) spectrum shown in Figure 2a (red: UV-Vis-NIR, blue: FTIR), $SnP_2S_6$ was found to be transparent from 0.54 μm to 8.5 μm.

To determine the complex refractive index, $\tilde{n} = n + ik$, prism coupling method (performed by Metricon Corp.), spectroscopic ellipsometry, and FTIR spectroscopy were employed. Since $SnP_2S_6$ belongs to the point group of 3, it is uniaxial with two distinct refractive indices: ordinary refractive index $\tilde{n}_o$ and extraordinary refractive index $\tilde{n}_e$. A GaP prism was used to measure $n_o$ and $n_e$ at the wavelengths of 0.633μm, 0.827μm and 1.55μm. The ellipsometric spectra collected on the (0001) surface of $SnP_2S_6$ were fitted to three Tauc Lorentz oscillators to extract $\tilde{n}_o$ from 0.200μm to 1.000μm (see Table S1 and Figure S2). The results agree well with the values measured by the prism coupling method. To obtain the ordinary optical constants in the spectral range of 1μm to 8.5μm, the interferometric method was used.[26] Knowing the thickness, *t*, of the sample and $n_o$ at the wavelength of the first peak in the transmittance (*T*) spectrum from Figure 2a, the order of interference, *m*, can be found using

$$m = \frac{2 \cdot n(\lambda_m) \cdot t}{\lambda_m}. \tag{1}$$

The refractive indices at the successive peaks can thus be determined by

$$n(\lambda_{m-i}) = \frac{(m-i) \cdot (\lambda_{m-i})}{2t}. \tag{2}$$

Since the energy range is below the bandgap and $n_o \gg k_o$, the reflectance *R* can be calculated using $R = \frac{(n-1)^2}{(n+1)^2}$. The absorption coefficient $\alpha$ can be estimated using $\alpha = -\frac{1}{t} \ln \frac{T}{(1-R)^2}$,[27] and thus $k_o$ can be found since $k_o = \frac{\alpha \lambda}{4\pi}$.

FP calculations have also been used to compute the electronic contributions to the frequency-dependent refractive index using the LDA and vdW structures. The experimental and calculated complex ordinary refractive index are shown in Figure 2b, demonstrating a very good agreement. The experimental $n_o$ and $n_e$ was then fitted to the Sellmeier equation:

$$n^2 = A + \frac{B}{\lambda^2 - C} + D \cdot \lambda^2 + E \cdot \lambda^4 \tag{3}$$

The parameters of the Sellmeier equations are shown in Table 1, and the real ordinary and extraordinary refractive indices are shown in Figure 2c. Note that $n_e$ beyond 1.55µm were obtained by extrapolating the Sellmeier equation to 6 µm based on the fact that it is non-absorbing up to 8.5 µm (Figure 2a).

**Table 1.** The parameters of Sellmeier equation for the linear optical properties of $SnP_2S_6$.

| $n$ | A | B(µm$^2$) | C (µm$^2$) | D(µm$^{-2}$) | E(µm$^{-4}$) |
|---|---|---|---|---|---|
| $n_o$ | 6.645 | 0.4121 | 0.06588 | -0.001321 | -6.716×10$^{-5}$ |
| $n_e$ | 4.551 | 0.2003 | -0.0268 | - | - |

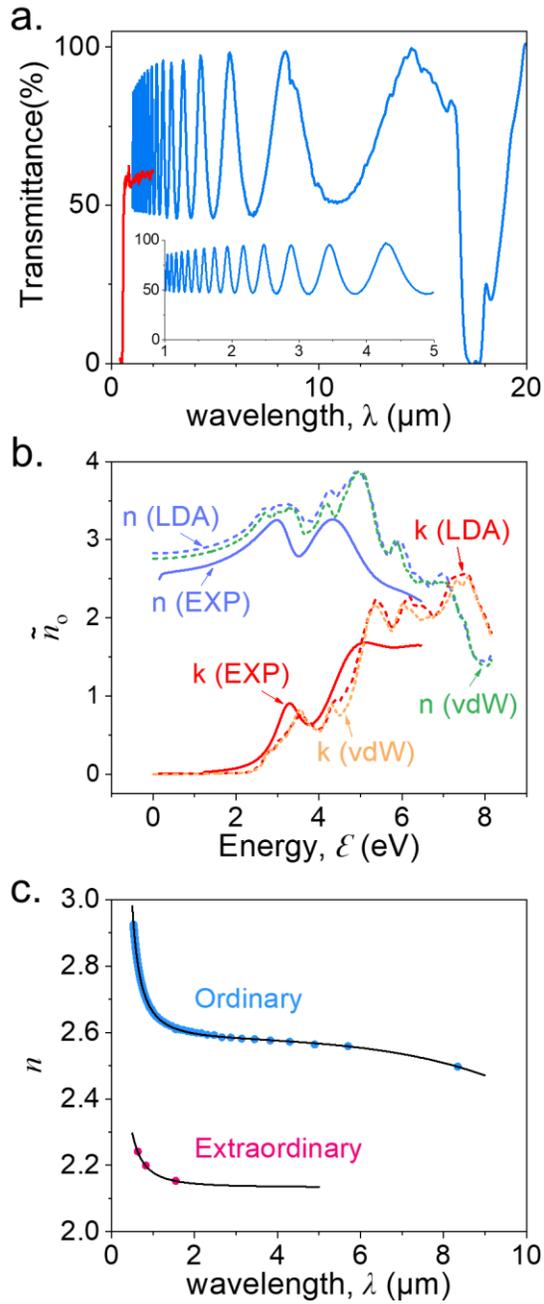

**Figure 2.** (a) Transmittance spectra of measured from 0.45μm to 20μm using UV-Vis-NIR spectroscopy (red) and FTIR spectroscopy (blue). The inset enlarges the FTIR spectrum from 1μm to 5μm for clarity. (b) The experimental complex ordinary refractive index of $SnP_2S_6$ compared with results from FP calculations using both LDA and vdW structures. (c) Ordinary and extraordinary refractive indices of $SnP_2S_6$ and the theoretical Sellmeier fit.

Next, we studied the second-order NLO susceptibility of $SnP_2S_6$. The induced nonlinear polarization $P_{2\omega}$ and the incoming electric field $E_\omega$ are related by the second-order optical susceptibility $P_{i,2\omega} \propto d_{ijk} E_{j,\omega} E_{k,\omega}$. The fundamental wavelength used was 1550nm and the resulting SHG wavelength was thus 775nm. This ensured that the SHG process occurred within the bandgap of 2.3eV and only involved virtual transitions,[28] and therefore minimized absorption and was non-resonant. SHG polarimetry was employed to study the SHG performance of $SnP_2S_6$, as shown in Figure 3a. At various incident angles, $\alpha$, the SHG responses were recorded when the linear polarized fundamental electric field $E_\omega=(E_\omega\cos(\psi), E_\omega\sin(\psi), 0)$, with regard to the laboratory coordinates (X, Y, Z), was rotated by an angle of $\psi$. By convention, the crystal physics axes ($Z_1$, $Z_2$, $Z_3$) are defined as $Z_1//[2\bar{1}\bar{1}0]$, $Z_2//[01\bar{1}0]$ and $Z_3//[0001]$.[29] Two sample orientations were used during the measurement, defined as O1: X = $Z_2$, Y = -$Z_1$, Z = $Z_3$, and O2: X = $Z_1$, Y = $Z_2$, Z = $Z_3$. The transmitted SHG field was then decomposed into *p*-polarized (∥) and *s*-polarized (⊥) by an analyzer and detected by a photo-multiplier tube. Figure 3b shows the quadratic dependence of SHG intensities and the incident power, confirming that the signal measured was generated by the second-order NLO effect.

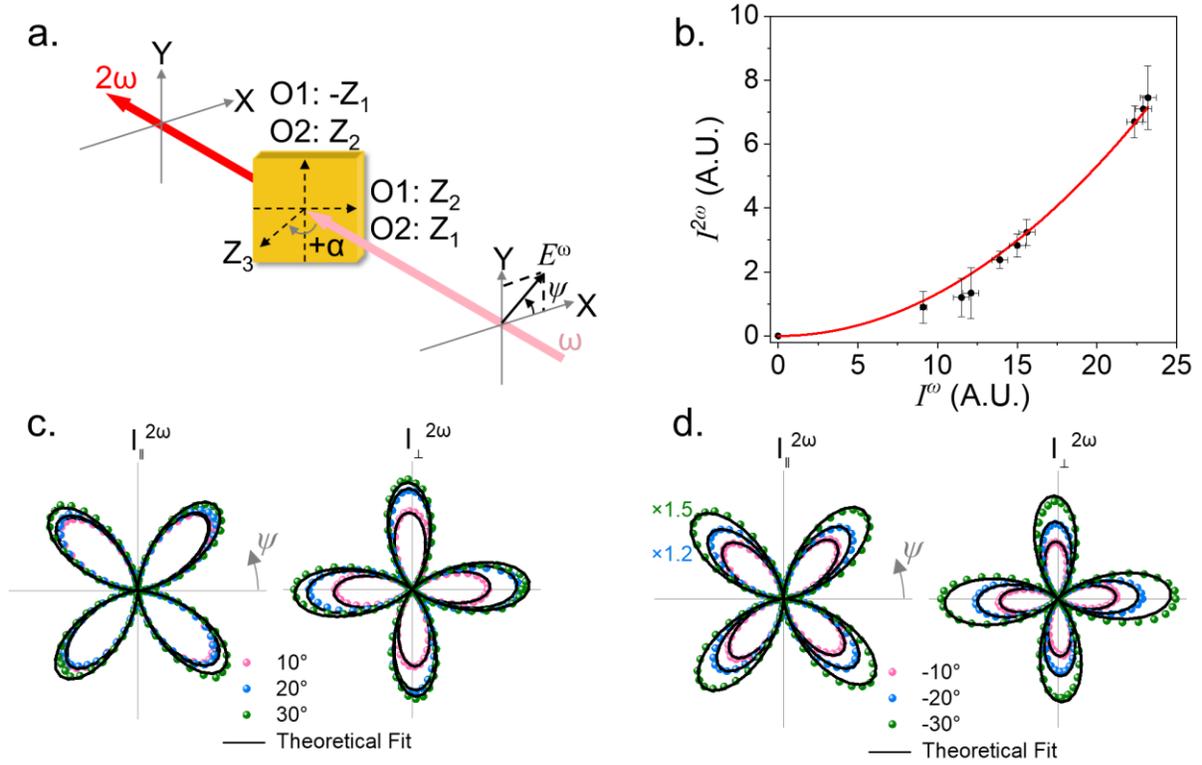

**Figure 3.** (a) Schematic of the SHG polarimetry geometry. (b) SHG power dependency of SnP$_2$S$_6$ crystal, confirming the detected response was from the SHG process. (c) Polar plots of *p*-polarized and *s*-polarized intensities for O1 at various incident angles. All sets of data were fit simultaneously to the point group 3.

For point group 3, the *d* tensor in Voigt notation is:

$$d = \begin{pmatrix} d_{11} & -d_{11} & 0 & d_{14} & d_{15} & -d_{22} \\ -d_{22} & d_{22} & 0 & d_{15} & -d_{14} & -d_{11} \\ d_{31} & d_{31} & d_{33} & 0 & 0 & 0 \end{pmatrix}. \qquad (4)$$

For O1, the $d_{\text{eff}}$ can be expressed as:

$$d_{eff,\parallel} = \cos\alpha_{2\omega}\,(d_{22}t_\parallel^2\cos^2\psi\cos^2\alpha_\omega - d_{22}t_\perp^2\sin^2\psi - d_{14}t_\perp t_\parallel \sin2\psi\sin\alpha_\omega$$

$$- d_{15}t_\parallel^2\cos^2\psi\sin2\alpha_\omega + d_{11}t_\perp t_\parallel \sin2\psi\cos\alpha_\omega)$$

$$- \sin\alpha_{2\omega}\,(d_{31}t_\parallel^2\cos^2\psi\cos^2\alpha_\omega + d_{31}t_\perp^2\sin^2\psi + d_{33}t_\parallel^2\cos^2\psi\sin^2\alpha_\omega)$$

$$d_{eff,\perp} = d_{11}t_\parallel^2\cos^2\psi\cos^2\alpha - d_{11}t_\perp^2\sin^2\psi - d_{15}t_\perp t_\parallel\sin2\psi\sin\alpha + d_{14}t_\parallel^2\cos^2\psi\sin2\alpha -$$
$$d_{22}t_\perp t_\parallel\sin2\psi\cos\alpha \tag{5}$$

where $t_\parallel$ and $t_\perp$ are the Fresnel transmission coefficients for the *p*- and *s*-polarized light, respectively. Expressions of $d_{eff}$ for O2 are provided in the Supporting Information (Equation S1). A *z*-cut LiNbO$_3$ crystal was used as a reference and measured under the same experimental conditions. The SHG coefficients of SnP$_2$S$_6$ were obtained by fitting the collected polar plots to an analytical model similar to the one developed by Herman and Hayden,[30] except three inhomogeneous waves were considered instead of one, and then compared with the reference crystal. Figure 3c, 3d and Figure S3 show the SHG polar plots and theoretical fit for O1 and O2, respectively. The extracted SHG coefficients are compared with the FP calculation, as listed in Table 2. The results show clear anisotropy and large nonlinear optical susceptibility, especially in one component $|d_{33}| \cong 53$pm/V at 1550nm fundamental light, which is higher than the benchmark infrared NLO crystals AgGaS$_2$ and AgGaSe$_2$. The large anisotropy, i.e. a near vanishing $d_{22}$, can be understood as follows. Each monolayer of the SnP$_2$S$_6$ lattice can be considered as applying small distortions to a higher-symmetry C$_{3v}$ structure. If these distortions are ignored, each monolayer acquires a C$_2$ rotation axis, along the vertical direction in Figure 1a upper panel. Since the distortions are small in the actual lattice, $d_{22}$ is near vanishing. Indeed, a FP calculation using a symmetrized C$_{3v}$ monolayer resulted in zero $d_{22}$. FP calculations were carried out using both the LDA and vdW structures to provide insight into the large SHG coefficients. The static SHG tensor $d_{ij}$ and average $<d_{ij}>$ (defined in ref. 31) were obtained with the LDA and vdW structures with scissor shifts.[31] The initial bandgaps are 1.11 and 0.97 eV with the LDA and vdW structures, respectively. Therefore, the applied scissor shifts are different (1.22 and 1.36 eV, respectively), leading to similar results in the end (Table S2). Band diagrams calculated using both LDA and vdW structures including scissor shifts are provided in Figure S4. Even though both structures are similar, using the vdW structure without including a scissor shift leads to a significant increase of <$d_{ij}$>, as shown in Table S3. Indeed, a larger scissor shift leads to smaller SHG coefficients. Overall, $d_{ij}$ is very sensitive to both the lattice parameters and the scissor shift.

The frequency dependence of $d_{ij}$ represented in Figure 4 depicts the calculated complex $d_{ij}$ coefficients for both structures at fundamental photon energy, $\mathcal{E}$, from 0 to 1.24eV, and Figure S5 shows the calculated $d_{ij}$ from 0 to 5.5eV. Overall, the agreement with the experiment is good, particularly the relative magnitudes of each component and the evolution with $\mathcal{E}$. The $d_{33}$ component shows the largest discrepancy; it is the most prone to being affected by vdW interactions. We find that a mere improvement of the structure (going from LDA to vdW) is, however, not sufficient to reconcile theory and experiment. Therefore, we hypothesize that vdW corrections going beyond the structure (i.e. also acting on the electronic density) are needed to improve $d_{33}$. We leave such investigation for future work.

In order to theoretically investigate the origin of the good NLO response of $SnP_2S_6$, one can first look at the parent structure $SnS_2$ (both formed of very similar $SnS_6$ octahedra), depicted in Figure S6. This structure is centrosymmetric and therefore has zero nonlinear response. Yet, the natural centrosymmetry of $SnS_2$ can easily be artificially broken by rotating one every two layers by 180° around X or Y as shown in Figure S7. This noncentrosymmetric $SnS_2$ is only 0.5 meV/atom higher in energy than its centrosymmetric counterpart, demonstrating the weakness of the vdW interactions between the layers forming the material. More importantly, it now presents a nonlinear response. Overall, the electronic structures of $SnP_2S_6$, centrosymmetric $SnS_2$ and noncentrosymmetric $SnS_2$ are very similar (Figure 1, Figure S1 and Figure S10-15). The most striking difference is the bandwidth of the conduction band, which is larger in $SnS_2$. This can be rationalized by the presence of the P dimers in SnP$_2$S$_6$, creating vacancies (see Figure 1a) that ultimately make the conduction band slightly less dispersive since it is constituted of Sn and S states. Therefore, the joint density of states (JDOS) is similar in all these systems, leading to similar dielectric constants (Table S5).[32] Additionally, computing the static SHG tensor $d_{ij}$ and average $<d_{ij}>$ for noncentrosymmetric SnS$_2$, we obtained a response only slightly smaller than SnP$_2$S$_6$ (Table S6), related to the slightly more dispersive conduction band and hence lower JDOS. We concluded that the SnS$_6$ octahedra present in both SnP$_2$S$_6$ and SnS$_2$ are the origin of the good optical properties of these materials. However, SnS$_2$ being centrosymmetric, it simply cannot present a nonlinear response.

The importance of the isolated conduction band can also be verified by examining the calculated $d_{ij}$ spectra. Here we take the calculated $d_{11}$ as an example (Figure S5a). According to a sum rule,[33]

$$Re\{\chi^{(2)}(0,0,0)\} = \frac{2}{\pi}\mathcal{P}\int \frac{Im\{\chi^{(2)}(-2\omega,\omega,\omega)\}}{\omega} d\omega \tag{6}$$

the majority of the positive contribution to non-resonant $d_{11}$ comes from lower-energy peaks in Im[$d_{11}$], e.g. the one near 2.4 eV in Figure S4. Comparing with the computed band structure in Figure S4, we find that one-photon vertical transitions at 2.4 eV are only possible between the valence band and the isolated conduction band. The above observation suggests that the large NLO response is due to this intermediate conduction band enabling a double resonance process involving the valence, intermediate, and upper conduction bands, similar to double resonance enhancement mechanisms known in asymmetric quantum wells.[34] A quantitative verification of this potential double resonance mechanism will be examined in future work. Note that our experimental measurements involve fundamental and SHG energies that are below the bandgap energy of ~2.3eV, and hence non-resonant. However, the proximity of the fundamental and SHG photon energies to the double resonant levels is responsible for our large SHG coefficients.

**Table 2**. Second-order NLO tensor elements of SnP$_2$S$_6$ from the experiment and FP at 1550nm. Numbers within parentheses are error bars.

| $d_{ij}$ at 1550nm | Experiment (pm/V) | FP (pm/V) |
|---|---|---|
| $d_{11}$ | ±22 (0.4) | ±32 |
| $d_{15}$ | ±1.7 (0.2) | ±2.4 |
| $d_{31}$ | ±0.48 (0.2) | ±2.4 |
| $d_{22}$ | ∓2.1 (0.04) | ∓3.0 |
| $d_{14}$ | ±0.33 (0.1) | ∓3.0 |
| $d_{33}$ | ∓53 (6.4) | ∓15 |

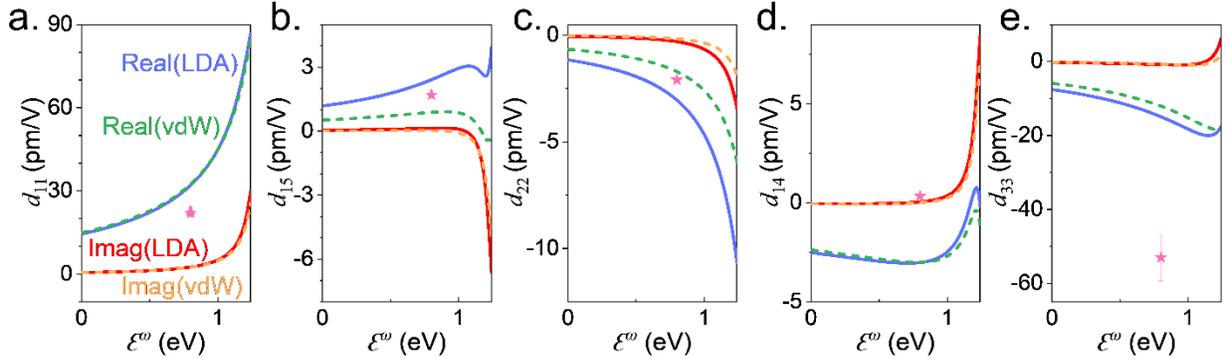

**Figure 4.** Complex $d_{ij}$ coefficients versus energy calculated from FP for the LDA (solid line) and vdW (dash line) structures. The real components are labeled in blue (green) and the imaginary components are labeled in red (orange) for the LDA (vdW) structure. The experimental values are highlighted with pink stars for comparison.

We also investigated the phase-matching conditions in SnP$_2$S$_6$. For applications of high-power laser systems, phase-matchability is one of the most critical criteria for a practical NLO material to have the most efficient SHG.[35,36] When an NLO material is phase-matched, the sum of the fundamental wavevectors is equal to the wavevector of the second harmonic radiation, or $n^\omega = n^{2\omega}$. However, due to the dispersion of the refractive indices, this can only be achieved when there is sufficient birefringence in the crystal. Since $SnP_2S_6$ is a negative uniaxial crystal ($n_o > n_e$), the Type-I phase-matching condition is $n_o^\omega = n_e^{2\omega}(\theta_m)$ and the Type-II phase-matching condition is $n_o^\omega + n_e^\omega(\theta_m) = 2n_e^{2\omega}(\theta_m)$. By calculating the phase-matching angles with the aforementioned Sellmeier equations, $SnP_2S_6$ was found able to achieve both Type I and Type II phase-matching conditions over a broad spectral range of 1-6μm, making it an excellent candidate for widely tunable lasers. Figure 5a and 5b show the Type I and Type II phase-matching angles ($\theta_m$) at wavelength, $\lambda$. The maximum $d_{eff}$ values at 1550nm wavelength are found to be 20.4 pm/V at $\phi$ =1.8° and 15.2 pm/V at $\phi$ =28.2° for Type I and Type II phase-matching respectively. Here, $\phi$ is the azimuthal angle.

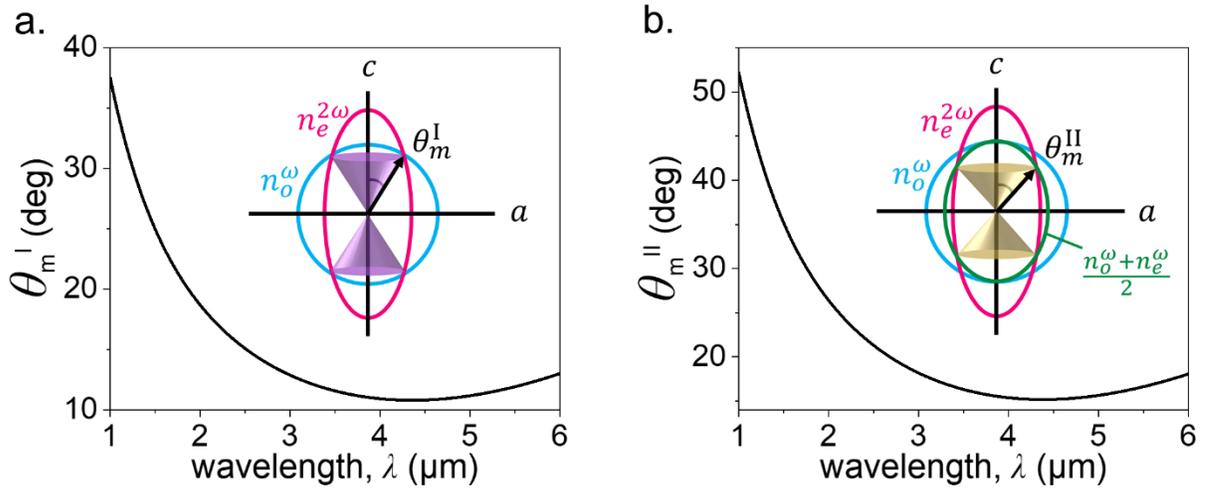

**Figure 5.** Type I (a) and Type II (b) phase-matching angles of $SnP_2S_6$ as a function of the fundamental wavelength, $\lambda$. The insets depict the relationship of the phase-matching angle and refractive indices.

To utilize a NLO crystal in actual applications, it is critical for the material to have large LIDT values as LIDT and NLO performance are competing requirements. The LIDT of $SnP_2S_6$ was assessed using a femtosecond laser system at 1550nm (1kHz, 100fs). The measurement was done on the (0001) plane of the crystal without polishing and coating. The incident power was gradually increased for each trial, and the surface was inspected using an optical microscope until any damage was observed. The LIDT value for $SnP_2S_6$ was found to be 350 GW cm$^{-2}$. This is more than three times greater than $ZnGeP_2$ (100 GW cm$^{-2}$ measured with 130fs pulses),[13] suggesting $SnP_2S_6$ can achieve a large SHG coefficient and laser damage threshold and thus be applied in high power infrared laser systems.

**Conclusion**

The linear and nonlinear optical properties of $SnP_2S_6$ bulk single crystal have been systematically studied. It exhibits both large non-resonant second-order nonlinear susceptibility of $d_{33}$=53pm V$^{-1}$ at 1550nm fundamental wavelength and a large LIDT value three times greater than $ZnGeP_2$. In addition, it can be both Type I and Type II phase-matched in a broad spectral range, and the effective $d$ coefficients at 1550nm are comparable to the state-of-art infrared NLO crystals $AgGaS_2$ and $AgGaSe_2$. The complete linear and SHG tensors measured experimentally and calculated by first principles theory are in excellent agreement. These remarkable properties suggest $SnP_2S_6$ could be a promising bulk material for next-generation infrared laser.

**Experimental Section**

Synthesis

A mixture of Sn powder (0.014mol, 1.7g), P powder (0.029mol, 0.89g) and S powder (0.086mol, 2.8g) was ground and sealed in an evacuated quartz tube. The quartz tube was heated from room temperature to 300°C in 5 hours and held at that temperature for 10 hours. After that, the source zone and the growth zone were heated to 400°C and 450°C respectively in 5 hours and held for 12 hours. Then the temperatures of the two zones were switched within 5 hours and maintained at that temperature for 8 days. Then the growth zone was cooled to 200°C in 5 hours while the source zone was maintained at 450°C. Finally, both zones were cooled to room temperature in 6 hours. The as-grown crystals were characterized by single-crystal x-ray diffraction (XRD) to demonstrate the phase purity and excellent crystallinity.

Spectroscopic Ellipsometry

The spectroscopic ellipsometry was carried out using a Woollam M-2000F spectroscopic ellipsometer with a focused beam. The ellipsometric spectra were collected on the (0001) surface of $SnP_2S_6$ from 0.200μm to 1.000μm at room temperature and fitted to three Tauc Lorentz oscillators to extract the complex ordinary refractive index. The parameters of each Tauc Lorentz oscillator include an amplitude $A$, broadening parameter $B$, central transition energy $E_0$ and Tauc gap $E_g$, as shown in Table S1.

SHG measurements

The SHG measurements were performed in transmission geometry. The 1550nm fundamental laser beam was generated by Coherent Libra Amplified Ti:Sapphire Laser (85fs, 2kHz) and focused on the sample surface. The *p*-polarized and *s*-polarized SHG intensities were recorded by a photo-multiplier tube while the fundamental beam was linearly polarized and rotated by an angle of φ with a half waveplate with respect to the lab X axis. A z-cut $LiNbO_3$ (MTI Corporation) was used as a reference.

First-principles calculations

First-principles (FP) calculations have been performed using density-functional theory (DFT) with ABINIT.[37–39] The wavefunctions were expanded on a plane-wave basis set with a kinetic energy cutoff of 40 Ha. Only valence electrons were explicitly taken into account adopting optimized norm-conserving Vanderbilt pseudopotentials from the PseudoDojo.[40,41] The exchange-correlation energy was modeled using the local-density approximation (LDA).[42] In order to reach accurate and trustable results for third-order energy derivatives, ground-state properties have been very precisely determined by sampling the Brillouin zone with a Monkhorst-Pack 16x16x16 $k$-point mesh and converging self-consistent field cycles with a criterion of $10^{-22}$ Ha on the total energy.[43,44] In all calculations, a scissor shift has been applied to the conduction band to match the experimental band gap of 2.3 eV.

The primitive (trigonal) unit cell, with the same orientation as the experimental one, has been fully relaxed until a maximum force of 2.5 meV/Å on each atom was reached. The unit-cell relaxation has also been performed including van der Waals (vdW) interactions through the DFT-D3 method.[45] Note that, in that case, the Perdew-Burke-Ernzerhof generalized-gradient approximation[46] had to be utilized for the exchange-correlation energy, since the implementation of DFT-D3 is not available with LDA. Then, this relaxed cell has also been used to compute NLO coefficients using LDA, as the implementation of static NLO coefficients is limited to LDA. However, since it does not depend explicitly on the electronic density, the DFT-D3 method only induces corrections to the structure or properties related to derivatives of atomic positions such as interatomic force constants. Properties related to electric-field derivatives are not directly affected by such vdW corrections but rather indirectly by the crystal structure change.[47]

The static dielectric and SHG tensors have been obtained using density-functional perturbation theory.[48] Third-order energy derivatives with respect to the electric field are obtained using the ($2n$+1) theorem,[49] where only the ground-state and first-order wavefunctions are needed. The frequency-dependent dielectric and SHG tensors have been computed within the independent-particle approximation with the Optic utility of ABINIT. Additional empty bands have been included, up to 20 eV above the valence band maximum, leading to a total of 64 empty bands. A broadening of 0.002 Ha (54 meV) has been used to smoothen the spectrum by avoiding divergences in the sum-over-states approach. The refractive index and extinction coefficient can easily be obtained from the dielectric constant following textbook equations. The average of the static SHG tensor, $\langle d_{ij} \rangle$, is obtained with ref. 31

$$\langle d_{ij} \rangle^2 = \frac{19}{105}\sum_i d_{iii}^2 + \frac{13}{105}\sum_{i \neq j} d_{iii} d_{ijj} + \frac{44}{105}\sum_{i \neq j} d_{iij}^2 + \frac{13}{105}\sum_{ijk,\text{cyclic}} \left( d_{iij} d_{jkk} + \frac{5}{7} d_{ijk}^2 \right),$$

where $d_{ijk}$ is the third rank SHG tensor and $d_{ij}$ its Voigt notation.[50]

The AbiPy python package has been used to automate the input generation and workflow and analyze the results.[51]

ASSOCIATED CONTENT

**Supporting Information**. The following files are available free of charge.

Additional details on the experiment and supporting figures and tables. (PDF)

AUTHOR INFORMATION

Corresponding Author

Venkatraman Gopalan - Department of Materials Science and Engineering, Pennsylvania State University, University Park, Pennsylvania, 16802, USA; email: vxg8@psu.edu

Notes

The authors declare no competing financial interest.

ACKNOWLEDGMENT

J.H. and V.G. acknowledge Air Force Office of Scientific Research Grant number FA9550-18-S-0003. S.H.L., and Z.Q., and Y.W. were supported by the National Science Foundation's (NSF) No. DMR-1539916 at the Pennsylvania State University Two-Dimensional Crystal Consortium Materials Innovation Platform (2DCC-MIP). R.Z., L.M., N.A. and V.G. were supported by the NSF Materials Research Science and Engineering Center for Nanoscale Science, DMR-2011839. G.-M.R. acknowledge financial support from F.R.S.-FNRS. Y.W. was also supported by start-up funds from the University of North Texas. H.W. and V.G. were supported as part of the Computational Materials Sciences Program funded by the U.S. Department of Energy, Office of Science, Basic Energy Sciences, under Award No. DE-SC0020145 and DE-SC0012375. We thank John Jackson from Metricon Corporation for making the prism coupling measurement. The present research benefited from computational resources made available on the Tier-1 supercomputer of the Fédération Wallonie-Bruxelles, infrastructure funded by the Walloon Region under Grant Agreement No. 1117545. Computational resources have also been provided by the supercomputing facilities of the Université catholique de Louvain (CISM/UCL) and the Consortium des Équipements de Calcul Intensif en Fédération Wallonie Bruxelles (CÉCI) funded by the Fond de la Recherche Scientifique de Belgique (F.R.S.-FNRS) under convention 2.5020.11 and by the Walloon Region.

# Supporting Information

# SnP$_2$S$_6$: A Promising Infrared Nonlinear Optical Crystal with Strong Non-Resonant Second Harmonic Generation and Phase-matchability

*Jingyang He[†], Seng Huat Lee[§,∥], Francesco Naccarato[‡], Guillaume Brunin[‡], Rui Zu[†], Yuanxi Wang[§,^], Leixin Miao[†], Huaiyu Wang[†], Nasim Alem[†], Geoffroy Hautier[‡,⊥], Gian-Marco Rignanese[‡], Zhiqiang Mao[§,∥], Venkatraman Gopalan[†,∥]\**

[†]Department of Materials Science and Engineering, Pennsylvania State University, University Park, Pennsylvania, 16802, USA

[§]2D Crystal Consortium, Materials Research Institute, Pennsylvania State University, University Park, Pennsylvania, 16802, USA

[∥]Department of Physics, Pennsylvania State University, University Park, Pennsylvania, 16802, USA

[‡]Institute of Condensed Matter and Nanosciences, UCLouvain, 1348 Louvain-la-Neuve, Belgium

[⊥]Thayer School of Engineering, Dartmouth College, Hanover, NH, 03755, USA

[^] Department of Physics, University of North Texas, Denton, Texas, 76203, USA

**Ultraviolet-Visible-near-IR Spectroscopy (UV-Vis-NIR) spectroscopy**

The UV-Vis-NIR spectroscopy measurements were carried out in the spectral range of 0.45μm to 2μm on the Perkin–Elmer Lambda 950 spectrometer with a 150-mm integrating sphere. A micro-focus lens and an iris were used during the measurement, and the beam size was ~2mm in diameter. A photo-multiplier detector and an InGaAs detector were used for UV-VIS range of 0.45μm to 0.86μm and NIR range of 0.86μm to 2μm.

**Fourier-Transform Infrared (FTIR) Spectroscopy**

FTIR measurements were performed using a Bruker Hyperion 3000 Microscope with a 15× objective lens at room temperature. The transmittance spectra were collected on the (0001) surface of $SnP_2S_6$ in the spectral range of 1μm to 20μm (10000 $cm^{-1}$ to 500 $cm^{-1}$) at normal incidence. Opus software was used for the data acquisition.

**Laser induced damage threshold (LIDT) measurements**

The laser induced damage threshold (LIDT) of $SnP_2S_6$ was measured at 1550nm using a laser beam (100fs, 1kHz) generated by Spectra-Physics Ti: sapphire pumped OPA-800C. The beam was focused on the (0001) surface of the crystal using a 10-cm lens. The fundamental power was gradually increased and recorded for each trial. The surface of the sample was examined using an optical microscope (Zeiss Axioplan 2) with 10x magnification until obvious damage was noted.

**Transmission Electron Microscopy (TEM)**

The TEM specimen was prepared with ThermoFisher Helios NanoLab Dual-beam focused ion beam (FIB) system. The lamella was created with 30 kV ion beam and the transferred using in-situ lift-out method to the TEM grid. The lamella was then thinned down to 300-400 nm with 30 kV ion beam. 5 kV ion beam was used to further thin down the specimen to electron transparency and 2kV ion beam was used to gently polish the surface to reduce amorphization and Ga implantation.

The annular dark-field (ADF-) STEM was performed in the ThermoFisher Titan$^3$ S/TEM equipped with double spherical aberration correctors. The operating voltage was 80 kV with the beam current of 30 pA and dwell time of 2 µs, and the probe convergence angle was 30 mrad. The STEM images are drift-corrected after acquisition using the non-linear drift-correction algorithm.[1]

**Expression of effective $d$ for O2: $X = Z_1$, $Y = Z_2$, $Z = Z_3$**

$$d_{eff,\parallel} = \cos\alpha_{2\omega}\left(d_{11}t_\parallel^2\cos^2\psi\cos^2\alpha_\omega - d_{11}t_\perp^2\sin^2\psi - d_{14}t_\perp t_\parallel\sin2\psi\sin\alpha_\omega - d_{15}t_\parallel^2\cos^2\psi\sin2\alpha_\omega\right.$$
$$\left. + d_{22}t_\perp t_\parallel\sin2\psi\cos\alpha_\omega\right)$$
$$- \sin\alpha_{2\omega}\left(d_{31}t_\parallel^2\cos^2\psi\cos^2\alpha_\omega + d_{31}t_\perp^2\sin^2\psi + d_{33}t_\parallel^2\cos^2\psi\sin^2\alpha_\omega\right) \quad (S1)$$

$$d_{eff,\perp} = -d_{22}t_\parallel^2\cos^2\psi\cos^2\alpha + d_{22}t_\perp^2\sin^2\psi - d_{15}t_\perp t_\parallel\sin2\psi\sin\alpha + d_{14}t_\parallel^2\cos^2\psi\sin2\alpha -$$
$$d_{11}t_\perp t_\parallel\sin2\psi\cos\alpha$$

**Comparison of SnP$_2$S$_6$ with SnS$_2$**

The centrosymmetric and non-centrosymmetric SnS$_2$ structures are represented in Figure S6 and S7, while Figure S8 shows more details about the SnP$_2$S$_6$ structure than in the main text. Figure S9 defines various metrics to compare the octahedra, that are then given in Table S4 for all three structures, including vdW interactions or not. The electronic band structure and joint density of states of all three systems are given in Figure S10-S15. The static dielectric constants are given in Table S5. In non-centrosymmetric SnS$_2$, the static SHG tensor is given by

$$d = \begin{pmatrix} 0 & 0 & 0 & 0 & d_{31} & 0 \\ 0 & 0 & 0 & d_{31} & 0 & 0 \\ d_{31} & d_{31} & d_{33} & 0 & 0 & 0 \end{pmatrix}. \quad (S2)$$

The independent components are given in Table S6.

**Supporting Figures**

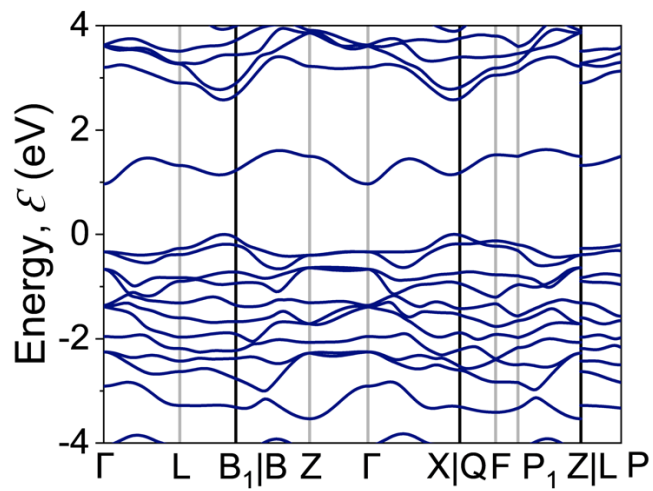

**Figure S1**. Band structure of SnP$_2$S$_6$ obtained from the vdW-relaxed structures.

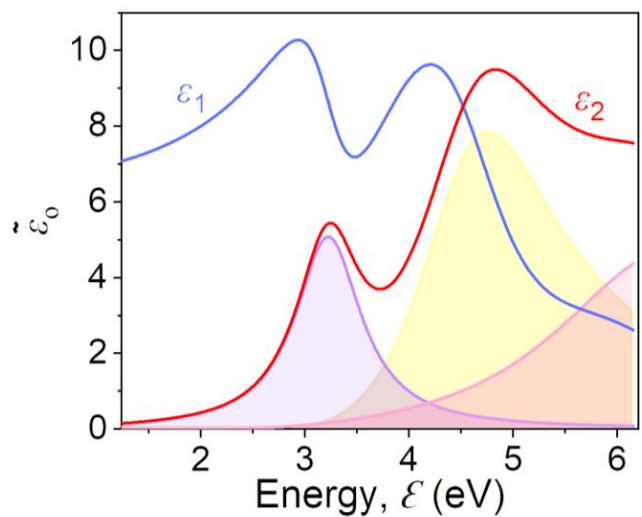

**Figure S2**. Complex ordinary dielectric constants of SnP$_2$S$_6$ extracted from ellipsometry, showing each Tauc-Lorentz oscillator.

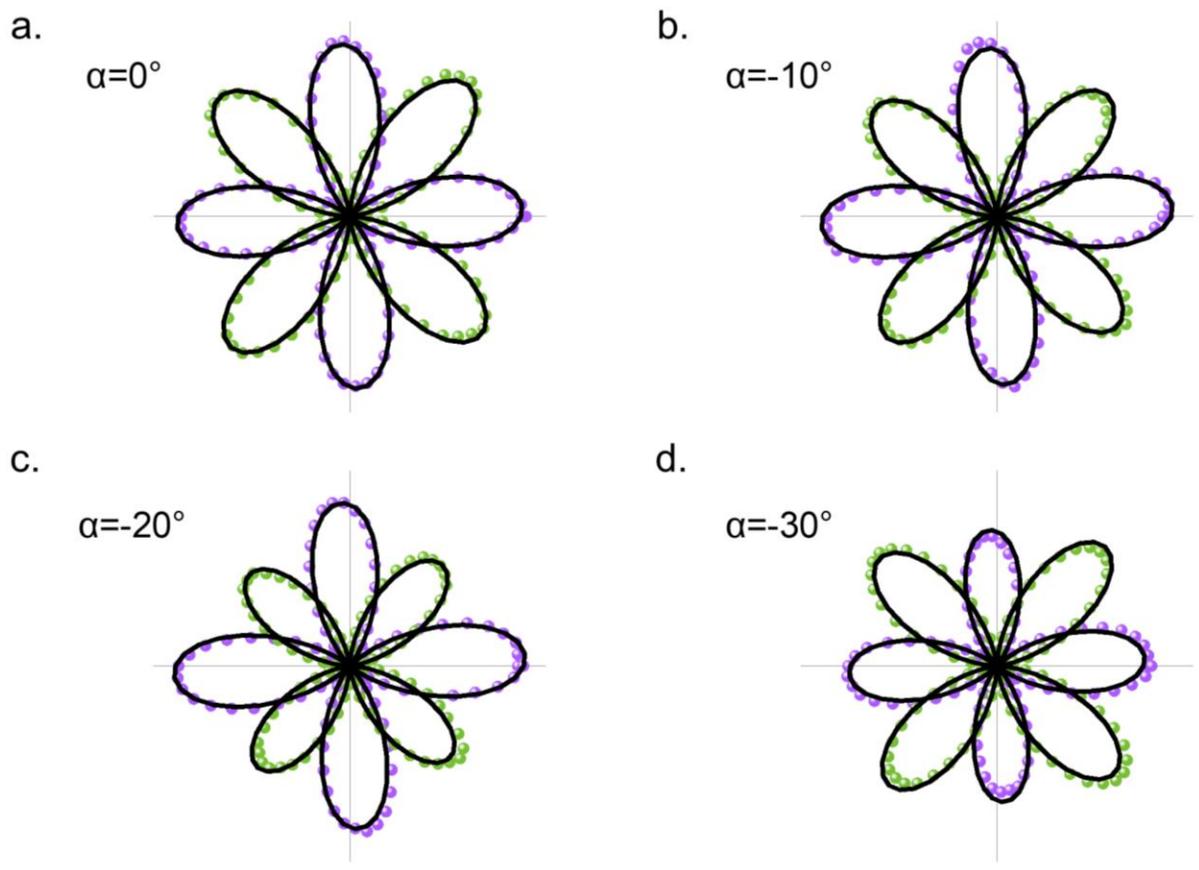

**Figure S3.** Polar plots of the *p*-polarized (purple) and *s*-polarized (green) intensities for orientation O2: X = $Z_1$, Y = $Z_2$, Z = $Z_3$ at various incident angles.

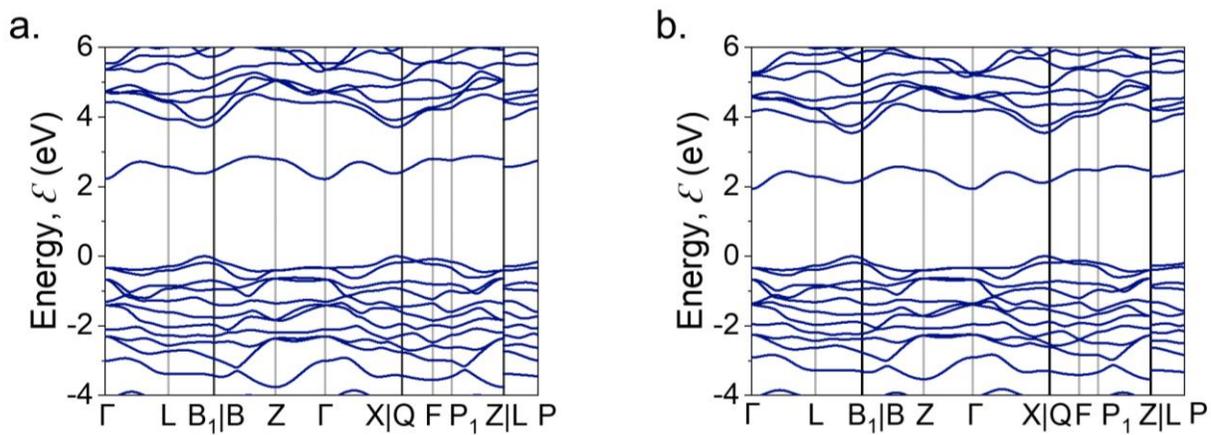

**Figure S4.** Band structure of $SnP_2S_6$ calculated with the LDA and vdW structures including scissor shifts of 1.22 and 1.36 eV, respectively.

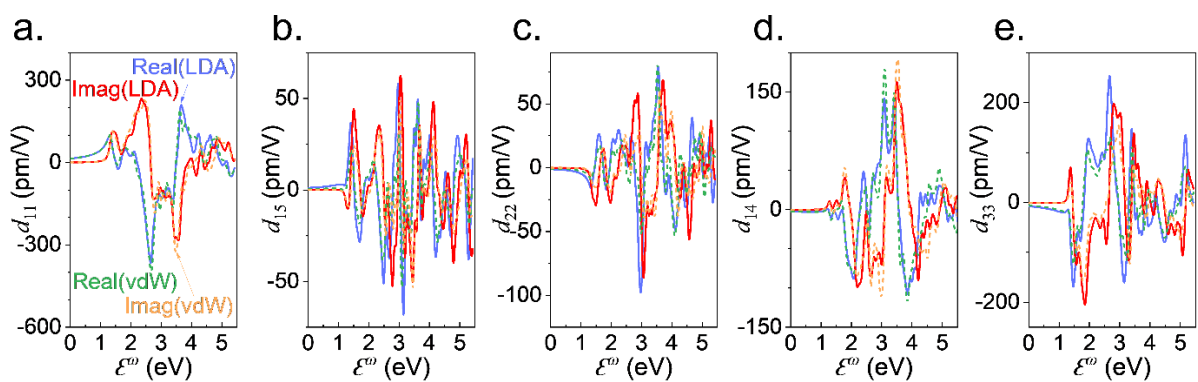

**Figure S5.** Complex $d_{ij}$ coefficients versus energy calculated for the LDA (solid line) and vdW structures (dash line).

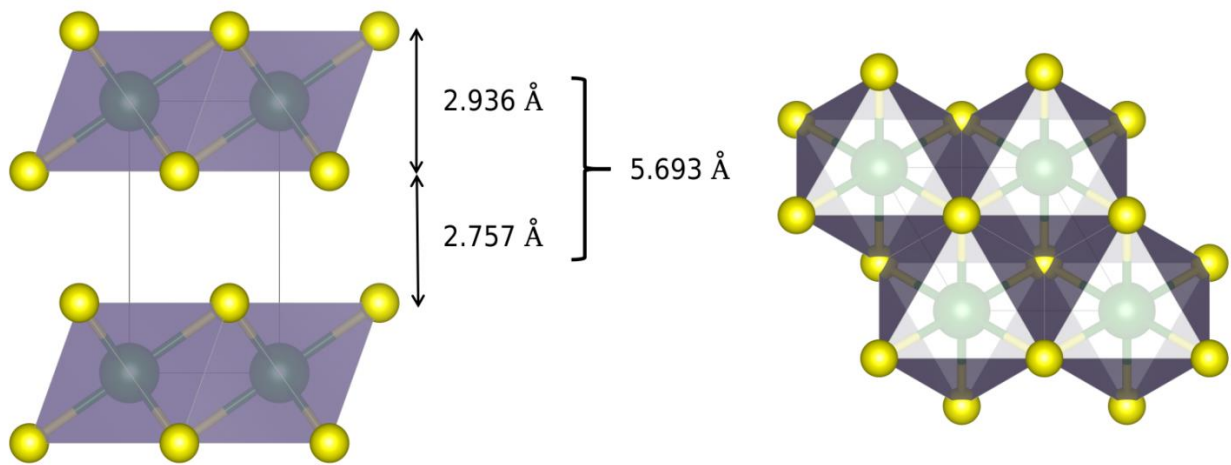

**Figure S6.** Crystal structure of centrosymmetric $SnS_2$. Including vdW interactions leads to an octahedra height of 2.965 Å and an interlayer distance of 2.940 Å.

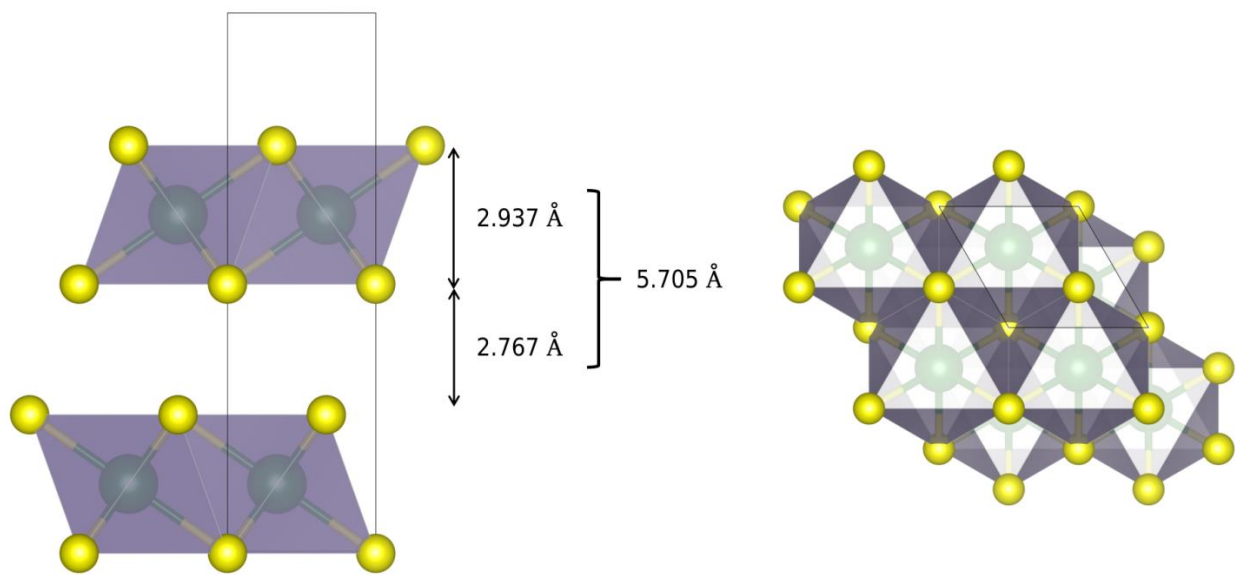

**Figure S7.** Crystal structure of non-centrosymmetric SnS$_2$. Including vdW interactions leads to an octahedra height of 2.966 Å and an interlayer distance of 2.949 Å.

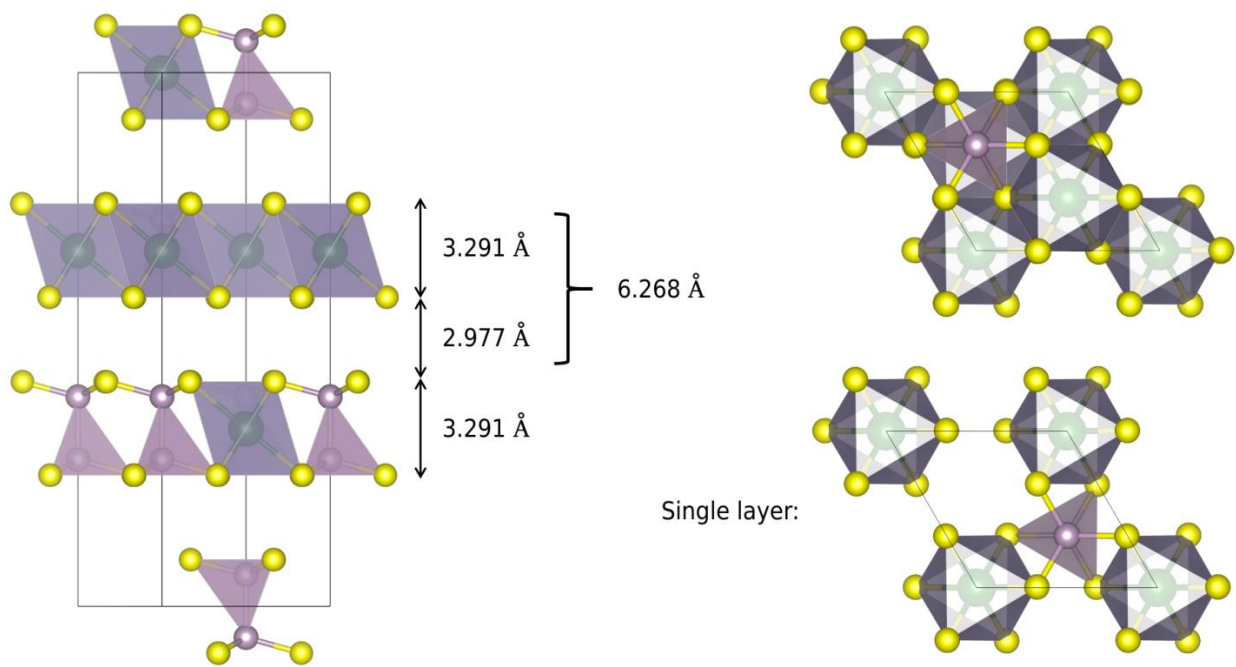

**Figure S8.** Crystal structure of SnP$_2$S$_6$. Including vdW interactions leads to an octahedra height of 3.319 Å and an interlayer distance of 3.168 Å.

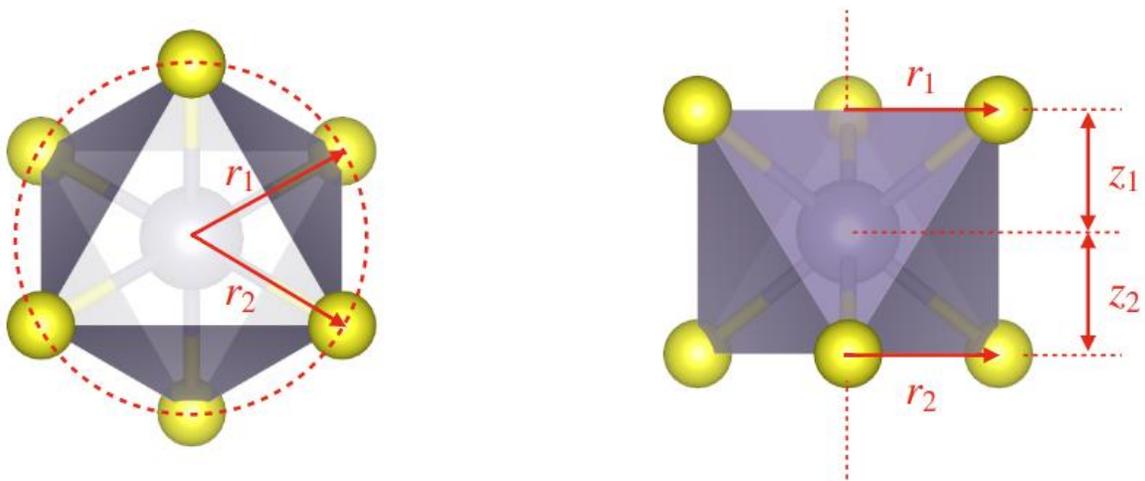

**Figure S9.** Definition of structural parameters for the SnS$_6$ octahedra.

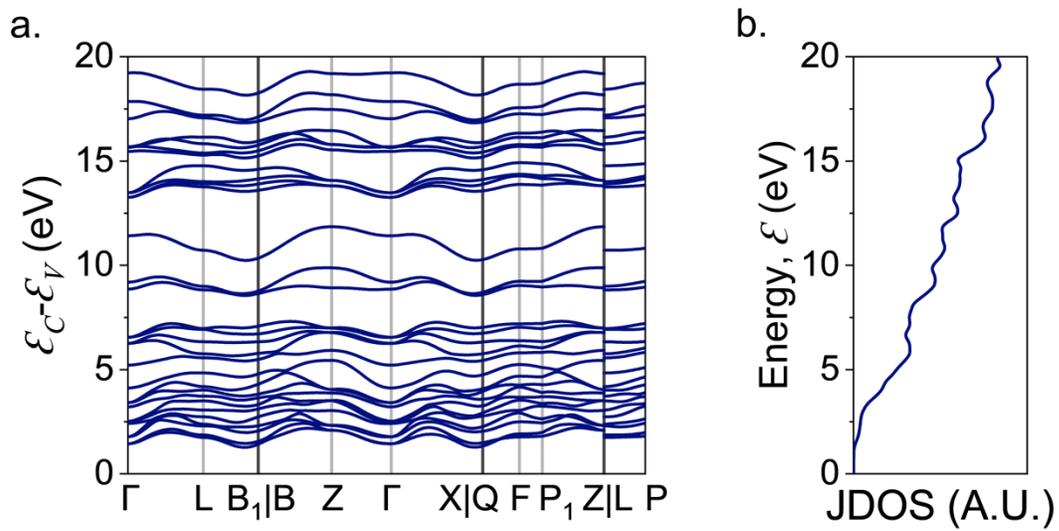

**Figure S10.** (a) valence to conduction band transition energies, and (b) joint density of sates (JDOS) of SnP$_2$S$_6$ for the LDA structure.

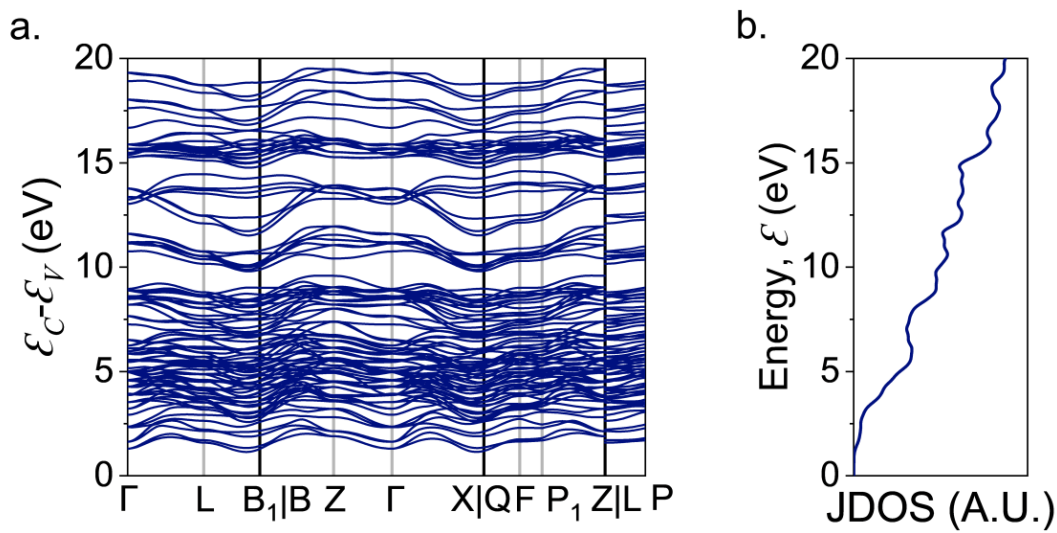

**Figure S11.** (a) valence to conduction band transition energies, and (b) joint density of sates (JDOS) of $SnP_2S_6$ for the vdW structure.

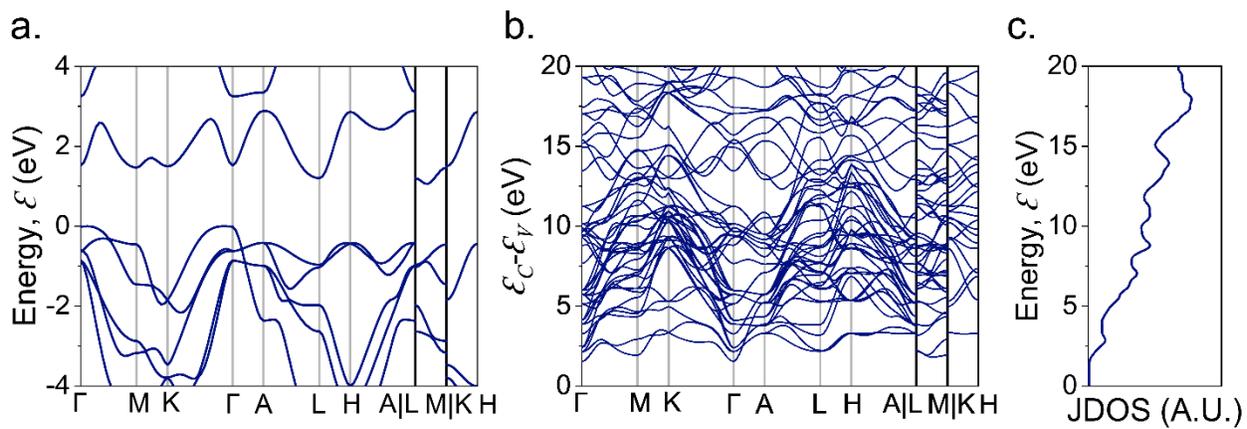

**Figure S12.** (a) Band structure, (b) valence to conduction band transition energies, and (c) joint density of sates (JDOS) of centrosymmetric $SnS_2$ for the LDA structure.

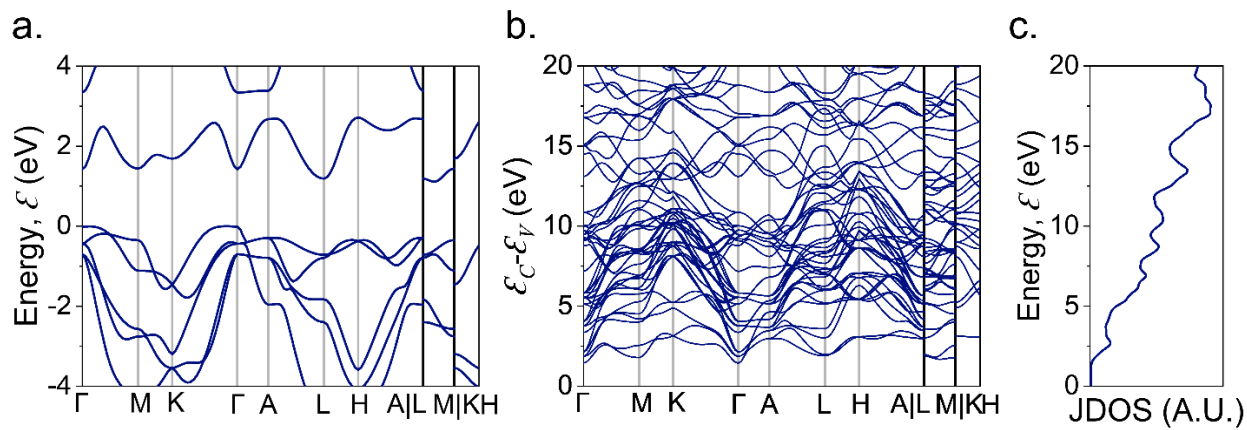

**Figure S13.** (a) Band structure, (b) valence to conduction band transition energies, and (c) joint density of sates (JDOS) of centrosymmetric SnS$_2$ for the vdW structure.

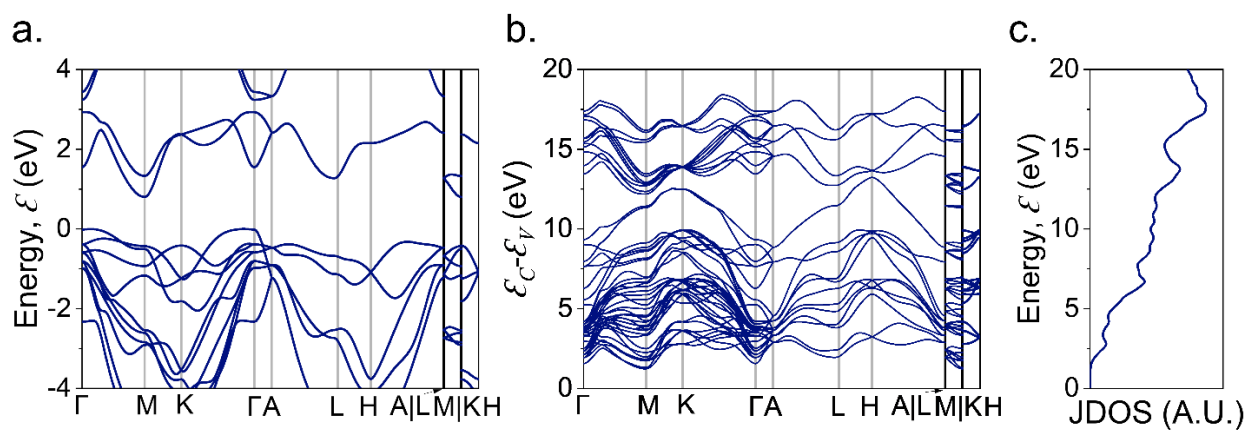

**Figure S14.** (a) Band structure, (b) valence to conduction band transition energies, and (c) joint density of sates (JDOS) of noncentrosymmetric SnS$_2$ for the LDA structure.

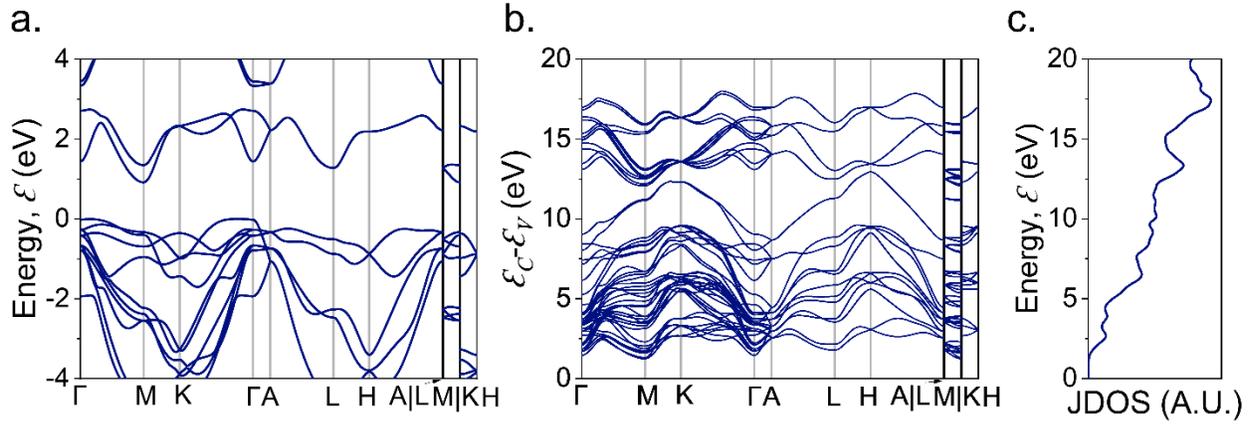

**Figure S15.** (a) Band structure, (b) valence to conduction band transition energies, and (c) joint density of sates (JDOS) of noncentrosymmetric $SnS_2$ for the vdW structure.

**Supporting Tables**

**Table S1**. The parameters of each Tauc-Lorentz oscillator for the linear optical properties (ordinary).

| m | $A_m^o$ | $B_m^o$ | $E_{0,m}^o$ | $E_{g,m}^o$ |
|---|---|---|---|---|
| 1 | 87.7380 | 1.664 | 4.611 | 2.867 |
| 2 | 3.9126 | 0.773 | 3.247 | 0 |
| 3 | 29.8659 | 2.821 | 6.543 | 2.167 |

**Table S2**. Static SHG tensor $d_{ij}$ (pm/V) obtained for the LDA and vdW structures. In both cases, a scissor shift has been applied to match the experimental bandgap.

| Static $d_{ij}$ | LDA (pm/V) | vdW (pm/V) |
|---|---|---|
| $d_{11}$ | ±21.7 | ±23.7 |
| $d_{15}$ | ±1.7 | ±0.9 |
| $d_{31}$ | ±1.7 | ±0.9 |
| $d_{22}$ | ∓2.2 | ∓1.4 |
| $d_{33}$ | ∓12.5 | ∓12.2 |
| <$d_{ij}$> | 15.9 | 17.1 |

**Table S3**. Static SHG tensor $d_{ij}$ (pm/V) obtained for the LDA and vdW structures. In both cases, no scissor shift has been applied.

| Static $d_{ij}$ | LDA (pm/V) | vdW (pm/V) |
|---|---|---|
| $d_{11}$ | ±55.1 | ±69.4 |
| $d_{15}$ | ±5.2 | ±2.0 |
| $d_{31}$ | ±5.2 | ±2.0 |
| $d_{22}$ | ∓6.9 | ∓4.6 |
| $d_{33}$ | ∓31.3 | ∓29.1 |
| <$d_{ij}$> | 40.4 | 49.4 |

**Table S4.** Structural parameters of the SnS$_6$ octahedra SnP$_2$S$_6$ and in centrosymmetric and non-centrosymmetric SnS$_2$. The numbers between parenthesis are those for the vdW structures.

|  | SnP$_2$S$_6$ | Centrosymmetric SnS$_2$ | Non-centrosymmetric SnS$_2$ |
|---|---|---|---|
| $r_1$ (Å) | 1.998 (2.022) | 2.099 (2.130) | 2.098 (2.129) |
| $r_2$ (Å) | 1.971 (2.007) | 2.099 (2.130) | 2.098 (2.129) |
| $z_1$ (Å) | 1.660 (1.668) | 1.468 (1.483) | 1.467 (1.482) |
| $z_2$ (Å) | 1.631 (1.651) | 1.468 (1.483) | 1.471 (1.484) |
| $r_1/r_2$ (%) | 101.4 (100.8) | 100 (100) | 100 (100) |
| $z_1/z_2$ (%) | 101.7 (101.0) | 100 (100) | 99.7 (99.9) |

**Table S5**. Static dielectric tensor of SnP$_2$S$_6$, and centrosymmetric and non-centrosymmetric SnS$_2$ calculated for the LDA structures. No scissor shift is applied.

| Static $\varepsilon_{ij}$ | SnP$_2$S$_6$ | Centrosymmetric SnS$_2$ | Non-centrosymmetric SnS$_2$ |
|---|---|---|---|
| $\varepsilon_{11}$ | 8.87 | 9.51 | 9.33 |
| $\varepsilon_{22}$ | 8.87 | 9.51 | 9.33 |
| $\varepsilon_{33}$ | 6.39 | 8.00 | 7.83 |

**Table S6**. Static SHG tensor $d_{ij}$ (pm/V) of non-centrosymmetric $SnS_2$ calculated for the LDA and vdW structures. No scissor shift is applied.

| Static $d_{ij}$ | LDA (pm/V) | vdW (pm/V) |
|---|---|---|
| $d_{31}$ | 37.5 | 30.0 |
| $d_{33}$ | 19.3 | 16.1 |
| $<d_{ij}>$ | 39.9 | 32.2 |